\documentclass{aa}  

\usepackage{graphicx}
\usepackage{xcolor}
\usepackage{txfonts}
\begin{document}

   \title{Star Cluster Ecology VIII. -- Dynamically Forming S-Type and $\alpha$Centauri-like Systems}

   \author{
        E. Hochart\inst{1} \and
        S. Portegies Zwart\inst{1}
    }
   \institute{
             \inst{1}Leiden Observatory, University of Leiden, 
             Niels Bohrweg 2, 2333 CA Leiden\\
             \email{spz@strw.leidenuniv.nl}
    }

   \date{Received 16/07/2026; accepted 07/10/2026}

 
\abstract
   {Approximately half the field stars in the local neighbourhood are part of binaries. Nevertheless, only $\sim15\%-20\%$ of detected exoplanets are around binary or higher-$N$ systems. Previous work has investigated the extent which binary systems modify planet formation or introduce dynamical instabilities, but few have considered the role of the cluster environment.}
   {To analyse the impact of dynamically formed binaries emerging naturally from various star cluster environments on already established planetary systems.}
   {Numerical simulations replicating the star forming region NGC 1333 are conducted. Three models are considered, each using identical cluster realisations and simulation parameters. The cluster itself is initially sub-virial and fractally distributed. Model~A assumes planetary systems are initially in mean motion resonance. In Model~B, the same planetary systems are moved out of resonance, but remain intrinsically stable. Finally, Model~C removes all planets from the simulation. Ten independent realisations are integrated per model until $30$~Myr.}
   {Dynamically formed binaries emerge at similar rates across models, indicating planets play little role in their formation. We also find a higher production rate of wide binaries compared to previous numerical studies which we attribute to the sub-virial and sub-clustered initial conditions. S-type and binary S-type planetary systems naturally emerge in our simulations, while P-type systems are rare. For Model~A, the fraction of S-type or binary S-type systems is $f_{\rm S}=0.02$ and $f_{\rm Bin,\, S}=0.05$, respectively. Rogue planets originating from a binary system are ejected with a substantially larger velocity than those around singletons. Planet-star collisions and planet capture events occur disproportionately within binary systems. Finally, several $\alpha$Centauri-like systems emerge in our simulations, suggesting that the nearest stellar system to the Sun may have a dynamical origin.}
   {}


   \maketitle
%
\section{Introduction}
    Stars are often found in pairs. Recent Gaia data finds that the binary fraction within $10$~pc of the Sun is $f_{\rm Bin}=0.274\pm0.023$ \citep{2021A&A...650A.201R}. Similarly, an earlier survey of the surrounding $25$~pc measured a fraction of $f_{\rm Bin}=0.33\pm0.02$ \citep{2010ApJS..190....1R}. This binary fraction itself seems to increase with the primary's mass, and the binary fraction of Solar-type stars in the field is $f_{\rm Bin.}\simeq0.46$ \citep{2010ApJS..190....1R}. Additionally, the fraction decreases in time, with young stars having a higher occupation fraction, the extreme being O-type stars (mass $m_*\gtrsim16$~M$_\odot$) with a near $\sim100\%$ primordial binary fraction \citep{2007A&A...474...77K, 2017ApJS..230...15M}. 
    
    These high rates contrast with observed exoplanetary systems. Indeed, \cite{2006ApJ...646..523R} find that the binary fraction for exoplanet hosts is $f_{\rm Bin.}\simeq0.23$, while \citet{2012A&A...542A..92R} reported an even lower value of $f_{\rm Bin}\simeq0.12$. Planetary systems hosted by a star part of a binary are generally S-type systems. S-type systems are those where planets orbit one of the stellar components. They account for $\sim95\%$ of observed planets in higher-$N$ systems. The remaining observed systems are P-type (circumbinary) systems, where planets orbit the binary stars' centre of mass.   
    
    The discrepancy between the observed large stellar multiplicity and the low proportion of planets whose host is part of a binary may partly result from observational effects. For instance, radial velocity measurements struggle to resolve tight binaries and are inherently biased against wide-orbit planets \citep{2005ApJS..159..141V, 2010ASSL..366...19E}. Conversely, false-positive companion detections may produce spurious binary populations \citep{2025arXiv251121817F}. Physical processes are also likely to contribute to the discrepancy. Dynamics between stellar components can lead to disk truncation depleting the reservoir needed to build planetary-mass objects \citep[e.g.,][]{1977ApJ...216..822P, 1996ApJ...467L..77A, 2008MNRAS.386..973P} or excite dynamical instabilities in planetary systems, leading to ejections and/or mergers \citep[e.g.,][]{1999AJ....117..621H, 2003ApJ...589..605W, 2021Ap&SS.366...53D, 2026MNRAS.545S2230Z}. This, however, is not completely detrimental, and planets-mass objects are still observed in protoplanetary disks within binary systems \citep[e.g.,][]{2008ApJ...681.1599M, 2013MNRAS.429..895P}.

    The evolution of binaries in star clusters constitutes a rich field in astronomy \citep[see i.e.][]{1972A&A....21..255A, 1975MNRAS.173..729H, 2024ApJ...975..207C}. In addition to processing primordial binaries \citep{2021MNRAS.501.4464C}, clustered environments can produce new binaries through dynamical interactions. Numerical studies have shown that roughly one wide binary, with semi-major axis $a\gtrsim10^{3}$~au, forms per cluster \citep{2009MNRAS.397.1577P, 2010MNRAS.404.1835K, 2011MNRAS.415.1179M}. Considering more compact systems, however, the formation rate and binary orbital parameters are sensitive to initial conditions, for instance, the degree of substructure present, the initial virial ratio and the cluster's density \citep{2023MNRAS.525.2907P, 2023MNRAS.518.4693G}.

    Previous work has studied the evolution of planetary systems part of a binary \citep{1999AJ....117..621H, 2007MNRAS.377L...1M, 2007ApJ...669.1298F, 2008ApJ...683.1063T, 2022MNRAS.515.2914E, 2026A&A...706A.121A}, however, the role of the cluster environment and the dynamical formation of the binary itself has received little attention. Here, we bridge this gap and simulate a star cluster, focusing on the emerging binary population and their circumstellar systems.
    
\section{Methodology}
    \subsection{The stars and planets}\label{Sec:Methods_Stars_Planets}
        The cluster simulated is modelled after the star-forming region NGC~1333, using initial conditions inferred from numerical simulations constrained by current observations. By comparing the spatial structure and degree of mass segregation in NGC~1333, \citet{2017MNRAS.468.4340P} find that this nearby \citep[$296$~pc;][]{2018A&A...619A.155S} young star-forming region is best reproduced by an initially fractal cluster with $150$~stars and fractal dimension $F_d=1.6$ \citep[following the prescription of][]{2004A&A...413..929G}, an initial half-mass radius $r_h=0.5$~pc and virial ratio $Q_{\rm vir}=0.3$ ($Q_{\rm vir}=0.5$ being virial equilibrium).
        
        Observationally, young clusters are often substructured \citep{2004MNRAS.348..589C,2009ApJ...696.2086S,2013ApJ...779..185K} and super-virial \citep{2019ApJ...870...32K,2024MNRAS.533..705W}. Indeed, using Gaia DR2 data, \citet{2019ApJ...870...32K} showed that $\gtrsim75\%$ of their $28$ targetted young clusters are super-virial. While unresolved binaries can inflate the measured velocity dispersion and thereby increase the inferred super-virial fraction  \citep{2010MNRAS.402.1750G}, it appears most stellar environments are super-virial. As such, our initially sub-virial cluster is not intended to provide a generic description of young cluster environments, but remains well motivated as it replicates the current structure of NGC~1333. Additionally, sub-virial environments do exist and have been observed, albeit at smaller rates \citep{2006A&A...445..979P,2015MNRAS.449..715W}. Lastly, although the adopted cluster has a small membership, the cluster membership distribution follows $dP/dN\propto N^{-2}$ \citep{2002MNRAS.336.1188K,2003ARA&A..41...57L}, meaning these small-$N$ clusters are considerably more common. Moreover, with fewer particles computational requirements are reduced.
        
        The initial mass for the $150$ stars are sampled from a \citet{2001MNRAS.322..231K} initial mass function with masses between $m_*\in[0.08,\, 30]\, {\rm M_\odot}$. We do not add a prescription for primordial binaries, but some do arise due to the randomised phase-space distribution of stars (cf. Section~\ref{Sec:Temporal}). Additionally, we do not model the embedded gas phase explicitly. Instead, calculations begin with a purely stellar system whose spatial and kinematic properties replicate the inferred early state of NGC~1333.
        
        \begin{figure}
            \centering
            \includegraphics[width=\columnwidth]{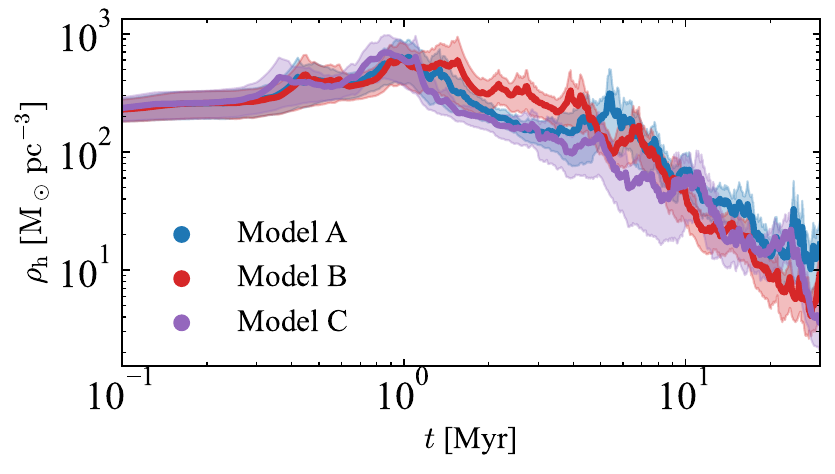}
            \caption{The evolution of the cluster's core density in time. As expected, identical stellar phase-space distributions between Model~A, Model~B and Model~C translate into little differences between models.}
            \label{Fig:cluster_density}
        \end{figure}
        We embed the cluster in a Milky Way–like potential \citep{2015ApJS..216...29B}, orbiting with Solar-like parameters \citep{2010MNRAS.403.1829S}. The external tidal field is coupled using the \texttt{BRIDGE} scheme \citep{2007PASJ...59.1095F} in \texttt{AMUSE} \citep{2013CoPhC.184..456P, 2013A&A...557A..84P, 10.1088/2514-3433/adfc3c}. We generate ten independent realisations, with the average cluster free-fall time being $\tau_{\rm ff}=0.62\pm0.02$~Myr. Figure~\ref{Fig:cluster_density} shows the evolution in the cluster density over time, computed following \citet{2003gmbp.book.....H},
        \begin{equation}
            \rho_{\rm core}=\frac{3(2^{2/3}-1)^{-3/2}}{4\pi}\frac{M_{\rm cluster}}{r_h^3},
        \end{equation}
        with $M_{\rm cluster}$ being the total stellar mass and $r_h$ the cluster half-mass radius. The cluster core has its density peak to values $\sim10^{3}$~M$_\odot$~pc$^{-3}$ before becoming an order unity by $\sim20$~Myr.
        
        Stars with masses $m_*\leq 2\,{\rm M_\odot}$ host planetary systems generated by the \texttt{PACE}\footnote{https://github.com/shuohuangGIT/PACE} population-synthesis framework \citep{2024A&A...689A.338H}.  Runs consist of about $145$ planetary systems, with planetary masses ranging between $3.15\times10^{-5}$~M$_{\rm Jup}$ and $2.33$~M$_{\rm Jup}$. Planets are initially on circular orbits and have a mutual inclination dispersion $\sigma_i=10^{-3}$~rads. Their initial semi-major axis ranges between an inner edge corresponding to one year orbits and a maximum of,
        \begin{equation}
            R_{\rm out}=117\left(\frac{M_*}{\rm M_\odot}\right)^{0.45}\,  {\rm au}. \label{Eqn:RDisk}
        \end{equation} 
        This follows from the mass-radius relations of protoplanetary disks found in numerical work and within the Ophiuchus region \citep{2010ApJ...723.1241A, 2020MNRAS.494.4130H, 2020ApJ...895..126H}.

    \subsection{The models}
        We consider three models: Model~A, B and C. All models use the same ten cluster realisations outlined above, but contain different planetary system architectures. That is, while planetary masses remain conserved, their orbital configurations differ.

\begin{figure}
            \centering
            \includegraphics[width=\columnwidth]{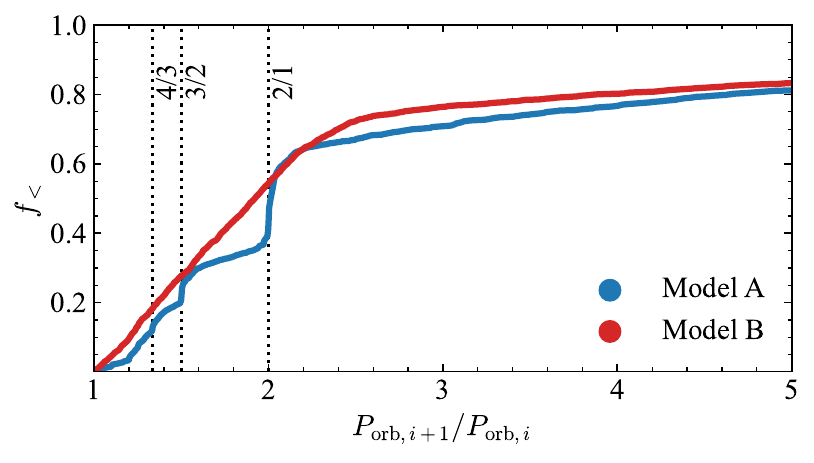}
            \caption{Cumulative distribution function of the initial ratio between a planet's orbit relative to its inner neighbour's.}
            \label{Fig:IC_MMR}
        \end{figure}

        In Model~A, when applicable, we initialise planetary systems in mean motion resonance (MMR) using the publicly available code\footnote{https://gitlab.strw.leidenuniv.nl/spz/Resonant\_chain}. This choice is motivated by theory suggesting MMR is a natural outcome of migration during planet formation \citep{2017MNRAS.470.1750I, 2022A&A...658A.170T, 2022MNRAS.511.3814H, 2023MNRAS.522..828H}. In Model~B, we break the MMR configuration by shifting a planet’s semi-major axis with a Gaussian-distributed random number. This translates into planets in Model~B having orbits within $0-10\%$ of their Model~A analogues. We ensure these systems remain stable at initialisation. Figure~\ref{Fig:IC_MMR} shows the cumulative distribution of planetary orbits relative to their inner neighbour. In Model~C, planets are excluded entirely. 
        
        With each model having ten  independent realisations, we perform $30$ runs, each being integrated until $t_{\rm end}=30$~Myr. This provides a more statistically robust analysis given the chaotic nature of $N>2$ body systems \citep{1891BuAsI...8...12P,1892mnmc.book.....P, 1986LNP...267..212D} while remaining computationally manageable.

    \subsection{The numerics}\label{Sec:Numerics}
        Integrations use the hybrid $N$-body code \texttt{Nemesis}\footnote{https://github.com/ErwanH29/nemesis} \citep{2019A&A...624A.120V, 2026A&A...709A..30H}, built using the \texttt{AMUSE} astronomical library. Given \texttt{AMUSE}'s `Noah's Ark' approach of having several codes available per physical regime, \texttt{Nemesis} provides user flexibility in tackling multi-scale and multi-physics problems in an astronomical setting. Here, the parametric code \texttt{SeBa} \citep{1996A&A...309..179P, 2012ascl.soft01003P} is used for stellar evolution.

        To accelerate integration, \texttt{Nemesis} separates fast-evolving systems such as planetary systems, binaries, and higher-order multiples, from the global environment. These fast-evolving systems (`children' or `child systems') are identified as the set of particles that are spatially separated by less than
        \begin{equation}
            R_{\rm par} = A\left(\frac{M_{\rm par}}{{\rm M}_\odot}\right)^{1/3},
        \end{equation}
        where $A=100$ and $M_{\rm par}$ is the total mass of the child system. Each child system gets integrated independently using a dedicated worker. Here, we use the symplectic code \texttt{Huayno} \citep{2012NewA...17..711P}. 
        
        When integrating the global environment, each child is replaced by a centre-of-mass representative particle (a 'parent'). This procedure removes tight orbits from the large-scale dynamics. The benefits of doing so is two-fold: Firstly, an optimised direct $N$-body code capable of handling large-$N$ systems is used when integrating the cluster. Secondly, fast-evolving systems such as planetary systems can still obtain high-precision solutions via symplectic codes. Isolated particles (i.e, rogue planets or single stars) are treated as self-representing parents. Here, the $4$th-Order Hermite integrator \texttt{Ph4} \citep{2012ASPC..453..129M} is used as the parent integrator.

        In \texttt{Nemesis}, local and global scales communicate with one another through a second-order kick-drift-kick scheme \citep{1967PhRv..159...98V}, whereby gravitational correction kicks are applied to the set of children and parents. This synchronisation occurs once per bridge step, set to $10^{3}$~yr. This is vital as it allows planetary systems to be influenced by the external potential, and vice versa (planets perturbing distant stars).
        
        As the cluster evolves, the children and parent population fluctuates. Children may partially or completely fragment, for instance, during a fly-by encounter, or when a planet is ejected. Such events are checked at every bridge step using a K-D tree connected-components search with linking length $1.5R_{\rm par}$. Particles connected within this distance remains a member of the same child. Any subset beyond the linking length is detached. If particles are isolated, they form self-representing parents. If they are near another particle, they form a new distinct child. 
        
        Conversely, parents may merge when the spatial separations between any two parent particles fall below their combined radii ($r_{ij}\leq (R_{{\rm par},\, i}+R_{{\rm par},\, j})$), in which case their associated children are combined into a new child system. The emerging child system is assigned a dedicated worker to enable high-accuracy solutions to its internal dynamics.

        \subsection{Identifying binaries and planetary systems}
        We identify bound systems through energy arguments. For binary stars, we iterate over the five nearest stellar neighbours, identifying a candidate binary as a system with eccentricity $e<1$ and an apocenter, $r_a=a(1+e)$, less than the minimum of either the system’s location-dependent Jacobi radius or the galactic tidal radius \citep{1987degc.book.....S, 2010MNRAS.401..977J}. If several pairs are identified, the binding energy for each candidate pair is computed using,
        \begin{equation}
            E_{\rm B} = -\frac{GMm}{2a}, \label{Eqn:Eb}
        \end{equation}
        with $G$ being the gravitational constant, $M$ and $m$ are the constituent masses, and $a$ the semi-major axis. The binary system is then identified as the pair with the lowest binding energy. We apply the same procedure to identify hierarchical triples, treating the centre-of-mass of each identified binary as a single object and searching for a bound tertiary companion. A hierarchical system has the additional requirement of needing to satisfy the long-term stability criterion of \citet{2001MNRAS.321..398M}.
        
        We adopt a similar approach when identifying planetary systems, except that we consider only the two nearest stars as potential hosts. A planet is S-type if it is most strongly bound to a single star, while a P-type occurs when the planet is most bound to the binary's centre of mass. We also apply this procedure to search for bound planet-planet binaries, but find none, consistent with \citet{2025NatAs...9..957P}.

\section{Results}
    \subsection{Time evolution of binaries}\label{Sec:Temporal}
\begin{figure}
            \centering
            \includegraphics[width=\columnwidth]{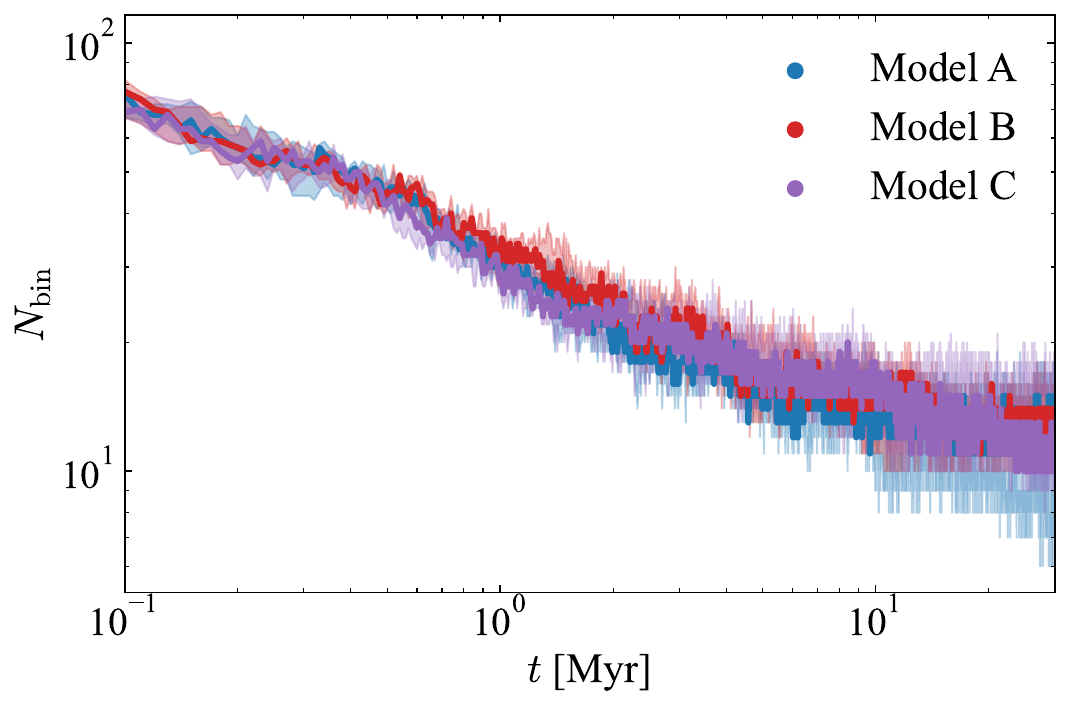}
            \caption{Median number of binaries over time. Filled region denotes the range in values within the $25\%$ and $75\%$ percentiles.}
            \label{fig:Nbin_time}
        \end{figure}

        Wide, weakly bound binaries will always arise due to statistical chance in a cluster \citep{2011MNRAS.415.1179M}. This is no different here even at the onset of simulation, however, the existence of this primordial population remains transient (Figure~\ref{fig:Nbin_time}). Model~A has $95^{+1}_{-21}$ primordial binaries with $0.04^{+0.01}_{-0.00}$ surviving the full $30$~Myr, meanwhile Model~B has $95^{+3}_{-19}$ with $0.04^{+0.03}_{-0.01}$ of them surviving until $30$~Myr. Finally, Model~C has $95^{+3}_{-19}$ primordial binaries with a survival rate of $0.05^{+0.00}_{-0.02}$ at $30$~Myr. 
        
\begin{figure}
            \centering
            \includegraphics[width=\columnwidth]{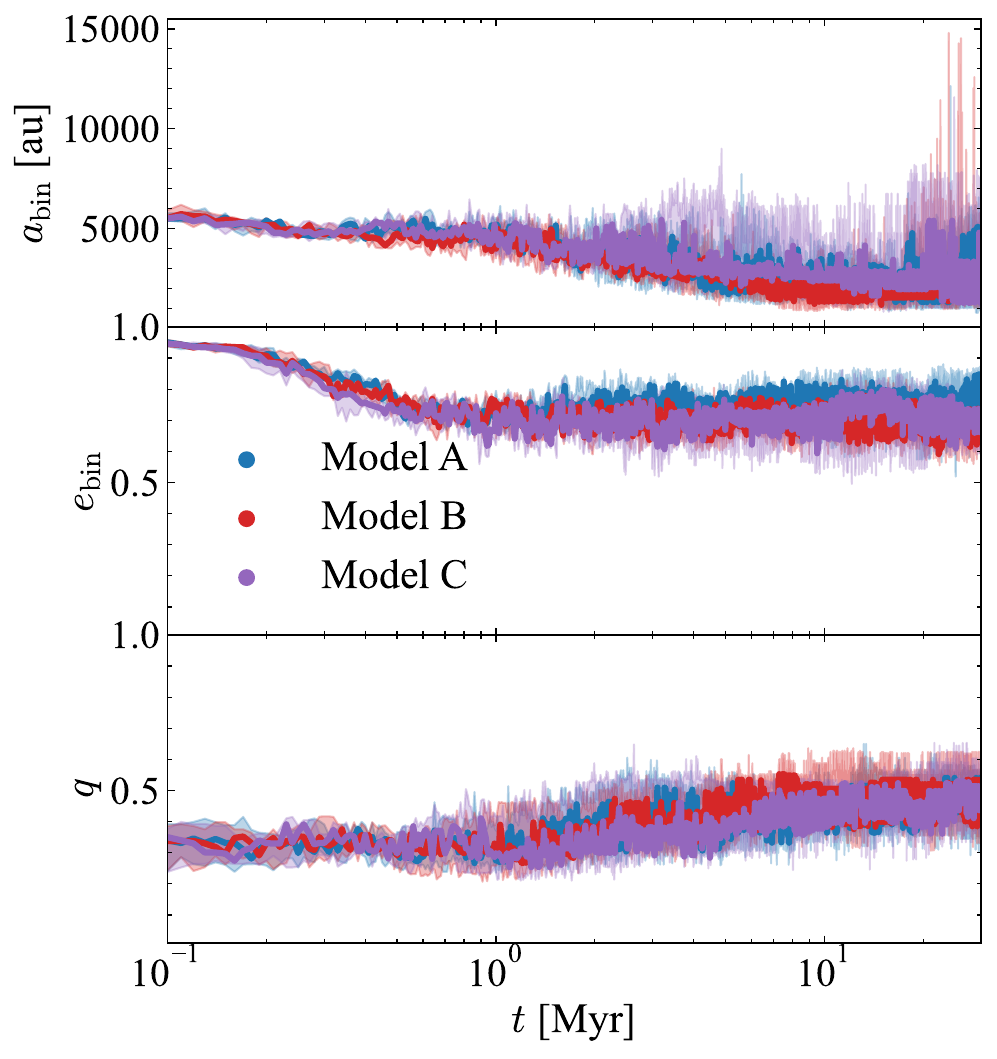}
            \caption{Median in binary orbital parameters over time. Filled region denotes the range in values within the $25\%$ and $75\%$ percentiles. Top, the binary semi-major axis. Middle, the binary eccentricity. Bottom, the binary mass-ratio.}
            \label{fig:Bin_OrbParam_time}
        \end{figure}

        The majority of the primordial binaries are soft, with the population being depleted to about a third of its initial value by $t\approx1$~Myr. This time coincides with the peak in the cluster density (Figure~\ref{Fig:cluster_density}), and also results in a heavy processing in their orbital elements (Figure~\ref{fig:Bin_OrbParam_time}). The weakly bound nature of the primordial binaries is also exemplified with the $e\approx1$ eccentricities they have at $t=0$~Myr. The distribution itself rapidly evolves through dynamical processing, and settles towards a roughly superthermal distribution, with a typical final eccentricity of $e\sim0.7$ (Table~\ref{Tab:BinStats}). 
        
        The binary mass ratio, $q$, slightly increases in time. Two mechanisms may contribute to this trend. At fixed total mass and semi-major axis, binaries with larger $q$ have a smaller cross-section and thus are less likely to undergo encounter-driven ionisation events. In addition, mass segregation concentrates massive stars in the dense cluster core. Here, frequent dynamical encounters and exchange interactions preferentially produce binaries containing more similarly massive components.
        
    \subsection{Binary demographics}\label{Sec:BinaryDemog}
        \begin{figure*}
                \centering                
                \includegraphics[width=.95\textwidth]{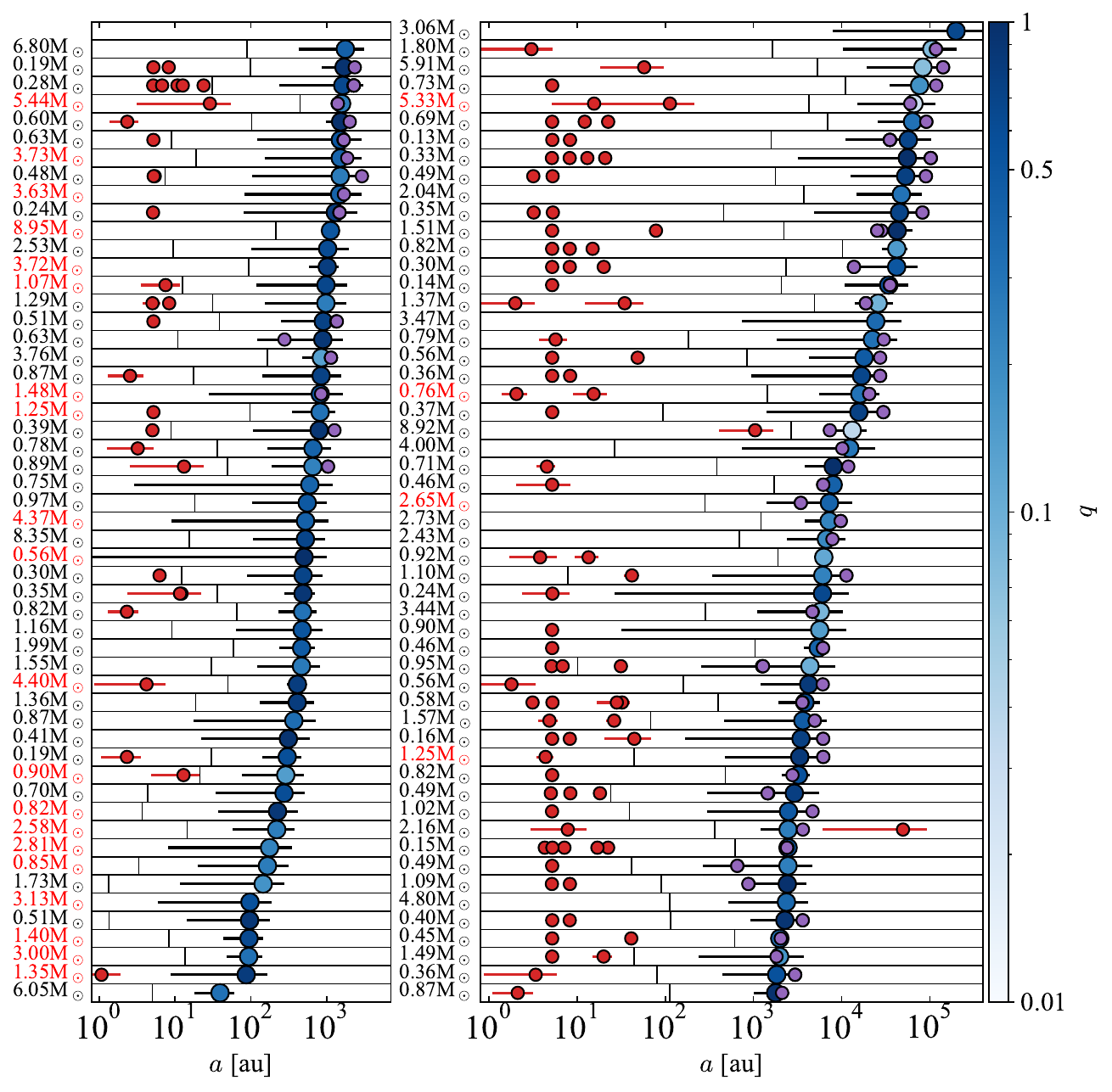}
                \caption{All dynamically formed binaries in Model~A. Red points denote planets bound to the primary star, purple points denote planets bound to the companion star. Large colour-graded points indicate the companion star, with colour representing the system's mass ratio. The mass labels in each row represent the primary's mass.  When coloured in red, the system is in a stable triple. For each system, planets orbiting the primary and companion stars are plotted at their orbital semi-major axes, $a$. Planets orbiting the companion are shifted relative to the binary system's semi-major axis. Horizontal lines show the extent of a body's orbit, extending between periastron and apoastron. Black vertical lines denote the system's region of stability. Any planet orbiting with a semi-major axis less than this is considered stable as per equation 1 of \citet{1999AJ....117..621H}. The underlying fit assumes coplanarity between the binary and planetary orbits and can deviate by up to $\sim11\%$ for high-eccentricity binaries with extreme mass ratios ($q\le0.1$ or $q\ge0.9$).}
                \label{Fig:ModelA_Overview}
        \end{figure*}
        \begin{table}
            \caption{Summary of results across all models ran. Upper and lower bound represent the interquartile range while the values themselves, the median. Row 1: Fraction of stars in binary systems. Row 2: Fraction of stars in binary but with no planets. Row 3: Fraction of stars in S-type systems. Row 4: Fraction of stars in binary S-type systems. Row 5: Median binary semi-major axis. Row 6: Median binary eccentricity. Row 7: Median binary mass ratio.}
            \label{Tab:BinStats}
            \centering 
                \begin{tabular}{c c c c}
                \hline\hline
                & Model~A & Model~B & Model~C\\
                \hline \vspace{-0.75em}\\
                $\langle f_{\rm Bin}\rangle$ & $0.09^{+0.01}_{-0.03}$ & $0.08^{+0.02}_{-0.01}$ & $0.08^{+0.03}_{-0.02}$ \vspace{0.2em}\\
                $\langle f_{\rm Bin.\, No}\rangle$ & $0.02^{+0.01}_{-0.01}$ & $0.03^{+0.00}_{-0.01}$ & -- \vspace{0.2em}\\
                $\langle f_{\rm S}\rangle$ & $0.02^{+0.01}_{-0.00}$ & $0.01^{+0.01}_{-0.00}$ & -- \vspace{0.2em}\\
                $\langle f_{\rm Bin.\, S}\rangle$ & $0.05^{+0.01}_{-0.04}$ & $0.04^{+0.01}_{-0.01}$ & -- \vspace{0.2em}\\
                $\langle a\rangle$ [au] & $2270^{+7930}_{-1640}$ & $2080^{+6150}_{-1480}$ & $2880^{+7520}_{-2090}$ \vspace{0.2em}\\
                $\langle e\rangle$ & $0.77^{+0.13}_{-0.24}$ & $0.74^{+0.16}_{-0.20}$ & $0.70^{+0.18}_{-0.24}$ \vspace{0.2em}\\
                $\langle q\rangle$ & $0.46^{+0.25}_{-0.20}$ & $0.51^{+0.23}_{-0.22}$ & $0.46^{+0.22}_{-0.25}$ \vspace{0.2em}\\ \hline
            \end{tabular}
        \end{table}
        
\begin{figure*}
                \centering                
                \includegraphics[width=.95\textwidth]{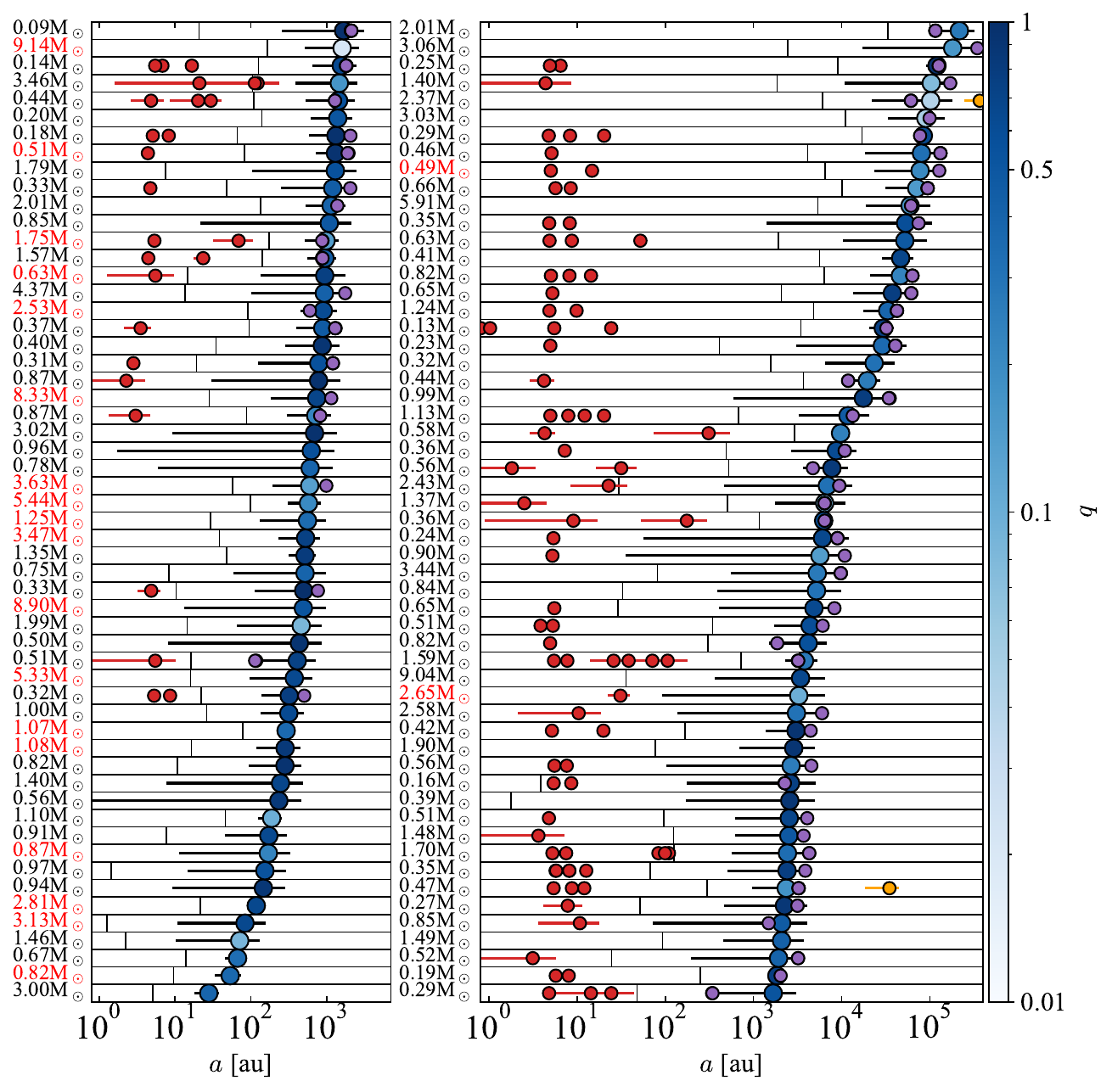}
                \caption{Same as Figure~\ref{Fig:ModelA_Overview} but for Model~B. A few P-type systems emerge, denoted by orange dots. Being so wide, however these will ionise at the next encounter.}
                \label{Fig:ModelB_Overview}
        \end{figure*}

        We define the fraction of stars part of a binary as,
        \begin{equation}
            f_{\rm Bin}\equiv\frac{N_B}{N_S+N_B+N_T},
        \end{equation} 
        were $N_B$, $N_S$ and $N_T$ denote the number of binaries, singles and triples respectively. The top four rows of Table~\ref{Tab:BinStats} summarises the types of binaries present after $30$~Myr between models.  A visual summary of the architecture of binaries in Model~A and Model~B is provided in Figure~\ref{Fig:ModelA_Overview} and Figure~\ref{Fig:ModelB_Overview} respectively.
        
        Model~A has slightly fewer binaries hosting no planets ($f_{\rm Bin. No}$) at the benefit of an enhanced binary S-type fraction ($f_{\rm Bin. S}$). Even so, the spread in results overlap with one another. For Model~B, two thirds of the planet-less binaries have at least one of the stellar component with mass $m_*>2$~M$_\odot$. Recall that stars with $m_*>2$~M$_\odot$ were initialised without planets (Section~\ref{Sec:Methods_Stars_Planets}). The fraction of binaries with semi-major axis $a>10^{3}$~au for Model~A and Model~B is $f_{\rm Bin, wide}=0.05^{+0.02}_{-0.02}$ and $0.05^{+0.01}_{-0.01}$ respectively.

\begin{figure*}
            \centering                
            \includegraphics[width=\textwidth]{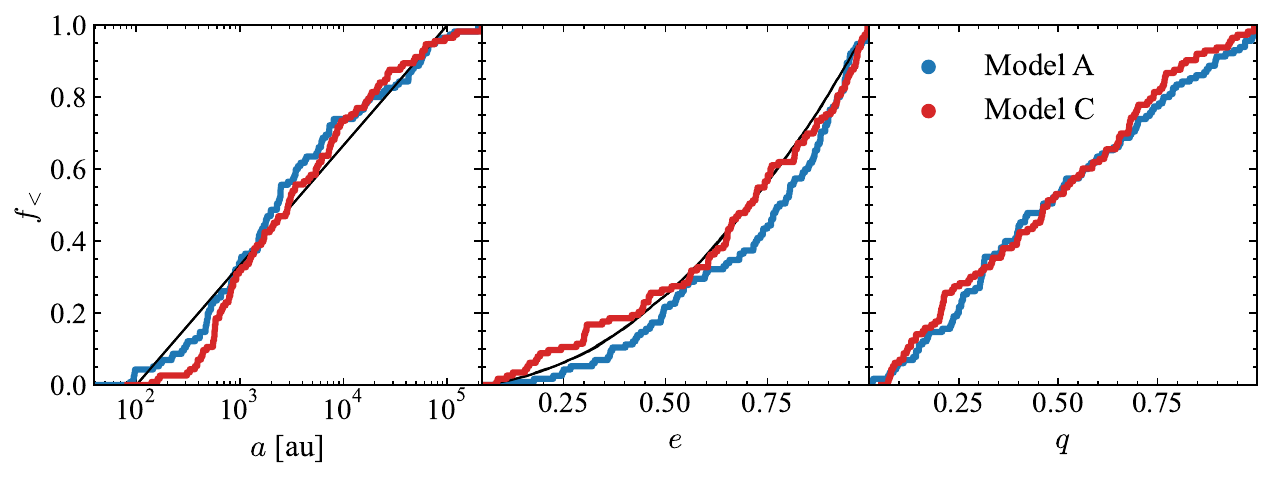}
            \caption{Cumulative distribution function of the semi-major axis (left), eccentricity (middle) and mass ratio (right) for all binaries formed. The solid black curve in the middle represents the thermal distribution.}
            \label{Fig:WithWithoutPlanets}
        \end{figure*}

        In most cases, the distribution of binary orbital parameters are also similar (Figure~\ref{Fig:WithWithoutPlanets}). Agreeing with theoretical expectations, dynamically formed binaries tend towards a thermal eccentricity distribution \citep{1919MNRAS..79..408J,1937AZh....14..207A,1975MNRAS.173..729H}. The semi-major axis of Model~A and B are lower than that of Model~C, which contains no planets, yet the spread in the data does overlap between all three models. A Kolmogorov-Smirnov test between Model~A and Model~C for all three variables have $p$-values of $p_a=0.34$, $p_e=0.1$ and $p_q=0.57$, meaning we cannot reject the null hypothesis that either distribution are drawn from different sets.
        
        \begin{figure}
            \centering
            \includegraphics[width=\columnwidth]{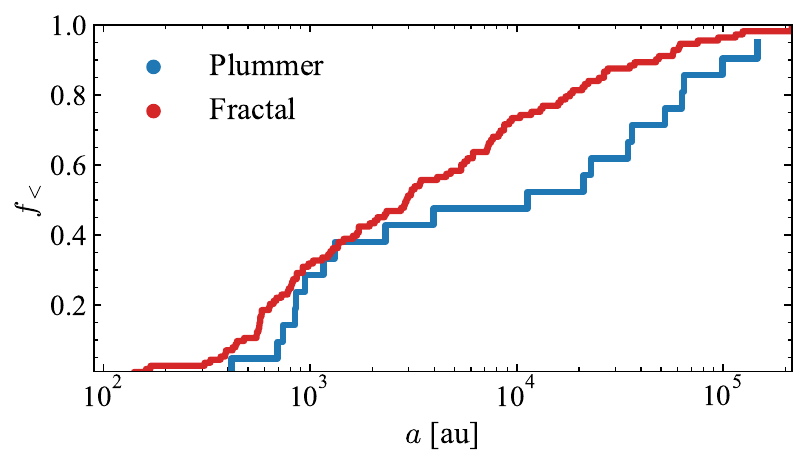}
            \caption{Cumulative distribution function of binary semi-major axis for the fractal (red, Model~C) and Plummer (blue) runs.}
            \label{fig:ModelA_vs_ModelC}
        \end{figure}
    
    \subsection{Collision events}\label{Sec:CollEvents}
        Model~A and Model~B experience $\langle N_{\rm coll}\rangle=8^{+1}_{-2}$ and $\langle N_{\rm coll}\rangle=7^{+2}_{-2}$ collisions respectively. Since there is a definite boundary between stellar-mass ($M_*>0.08$~M$_\odot$) and planetary-mass object ($m_p\leq2.33$~M$_{\rm Jup}$, or, equivalently  $\leq2.2\times10^{-3}~$M$_\odot$), we can distinguish the collision types. 
        
        For Model~A, $f_{\rm PP}=0.13^{+0.15}_{-0.02}$ of all collision events are planet-planet collisions, while the remainder are planet-star collisions. For Model~B, all collisions are planet-star collisions. The reduced number of planet-planet collisions occurring in Model~B is likely due to planetary systems being less compact.

        The quoted values include collisions across all systems, not only binaries. Despite this, collisions occur preferentially in binaries. For Models A and B, the fraction of collisions occurring in systems identified as binaries at $t=30$~Myr is $f=0.31^{+0.17}_{-0.16}$ and $f=0.23^{+0.17}_{-0.11}$, respectively. Accounting for systems that were a binary at collision (with age $>0.1$~Myr) but have been ionised by $t=30$~Myr, these fractions increase to $f=0.50^{+0.13}_{-0.00}$ and $f=0.53^{+0.08}_{-0.23}$. In either case, this fraction exceeds the respective model's binary fraction $f_{\rm Bin}\approx0.09$ (Table~\ref{Tab:BinStats}).

    \subsection{Capture events}\label{Sec:CaptEvents}
        Both models exhibit planetary capture events. Across all systems, $\langle N_{\rm cap}\rangle = 9.5^{+3.5}_{-0.5}$ and $\langle N_{\rm cap}\rangle = 9.5^{+2.3}_{-2.3}$ planets are captured per run for Models A and B respectively. When restricting the sample to binaries still present at $30$~Myr, these values decrease to $\langle N_{\rm cap}\rangle = 2.5^{+1.3}_{-1.5}$ and $\langle N_{\rm cap}\rangle = 2.0^{+1.8}_{-1.8}$, respectively. Overall, $f \approx 0.26$ and $f \approx 0.21$ of all captured planets reside in binaries.
        
        When accounting for systems that are no longer a binary at $t=30$~Myr, but that belonged to a binary with age $>0.1$~Myr during planet capture, the results for Model~B remains unchanged while for Model~A they increase to $\langle N_{\rm cap}\rangle = 7.0^{+1.0}_{-2.8}$. This raises the corresponding fraction to $f \approx 0.78$, indicating that captured planets in Model~A are strongly associated with present-day or former binary systems.

    \section{Discussion}
        \subsection{Impact of planetary system architecture}\label{Sec:System_Architecture}

            Regarding system architecture of S-type planetary systems, the increased resilience of planetary systems in MMR as found in \citet{2025A&A...700A.108M} is not obvious here when considering $f_{\rm Bin, No}$ (Table~\ref{Tab:BinStats}) nor in the number of collision events. This is partly attributed to not all planetary systems being in MMR (see Figure~\ref{Fig:IC_MMR}). However, to fully isolate the effect of compactness from that of MMR, we performed additional runs. These runs use the exact same planetary systems as Model~A, but randomise their initial orbital phase and argument of pericenter. Hereafer we refer to these runs as Model~D.
            
            Across the ten Model~D simulations, we find a median number of collision events of $\langle N_{\rm coll}\rangle = 14^{+1}_{-3}$, nearly twice the $\langle N_{\rm coll}\rangle = 8^{+1}_{-2}$ found for Model~A. This suggests that the higher collision rate in Model~A relative to Model~B is driven by its more compact planetary architectures. In addition, while the number of planet-star collisions is comparable across Models~A, B, and D, the number of planet-planet collisions differs strongly between models. Model~D yields $\langle N_{\rm coll,pp}\rangle = 7.0^{+2.0}_{-2.0}$, compared with $\langle N_{\rm coll,pp}\rangle = 1.0^{+0.4}_{-0.3}$ for Model~A and $\langle N_{\rm coll,pp}\rangle = 0.0^{+0.0}_{-0.0}$ for Model~B. The lower number of planet-planet collisions in Model~A relative to Model~D suggests that MMR can enhance the long-term stability of compact planetary systems.
            
            The increased resilience in a MMR configuration is also observed by analysing the complete rogue planet population. Model~A produces $\langle N_{\rm rogue}\rangle=77^{+7}_{-12}$ rogue planets per run. This translates to $f_{\rm rogue}=0.20^{+0.03}_{-0.02}$ of all planets being rogue. This is lower than that of Model~B ($\langle N_{\rm rogue}\rangle=105^{+23}_{-8}$, $f_{\rm rogue}=0.28^{+0.05}_{-0.02}$) and Model~D ($\langle N_{\rm rogue}\rangle=85^{+3}_{-4}$, $f_{\rm rogue}=0.23^{+0.00}_{-0.02}$). Taken together, these results indicate that while MMR helps stabilise systems, it is not so resilient to overcome consistent perturbations over secular time-scales.


        \subsection{Impact of cluster environment}\label{Sec:Env}
            \subsubsection{Wide-binary formation rates}\label{Sec:Wide_Bin_Forms}
            If fractal clusters are representative of young star-forming regions, they represent a viable channel for producing the observed wide-binary population ($a\gtrsim10^{3}$~au). Observations of the surrounding 100~pc find binary fractions of $f_{\rm Bin,F}=0.12$, $f_{\rm Bin,G}=0.12$, and $f_{\rm Bin,K}=0.10$ for F-, G-, and K-type stars \citep{2021A&A...649A...6G} (see also \citealt{2016ApJ...820L...2L,2019MNRAS.482L.139E}). Contrastingly, our simulations produce higher rates, with $f_{\rm Bin,F}=0.20^{+0.30}_{-0.15}$, $f_{\rm Bin,G}=0.28^{+0.38}_{-0.12}$, and $f_{\rm Bin,K}=0.08^{+0.09}_{-0.04}$. 
            
            The rates found here are also greater than those reported in previous work \citep{2009MNRAS.397.1577P,2010MNRAS.404.1835K,2011MNRAS.415.1179M,2011MNRAS.418.2565P,2017MNRAS.468.3461T}. This increased yield is particularly pertinent for the very wide population ($a>10^{4}$~au), although the small number of systems warrants caution. Models~A and B produce $\langle N_{\rm wide}\rangle=3.0^{+1.0}_{-2.0}$ and $\langle N_{\rm wide}\rangle=2.5^{+1.3}_{-1.3}$ per run, respectively, amounting to $f_{\rm wide}\approx0.02$. In contrast, across densities of $\rho=50-5\times10^{5}$~M$_\odot$~pc$^{-3}$, \citet{2009MNRAS.397.1577P} find that no wide binaries form in virialised Plummer models. Similarly, \citet{2011MNRAS.415.1179M} predict approximately one wide binary per cluster across densities of $\rho=10^{4}-10^{6}$~M$_\odot$~pc$^{-3}$. Both studies conclude that the very wide-binary formation rate is largely independent of cluster properties.
            
            \subsubsection{Influence of the initial cluster structure}
            To investigate the origin of the discrepancy, we perform an additional set of simulations, hereafter Model~E, using the same ten realisations of stellar properties but redistributing the stars into virialised Plummer profiles. This allows us to isolate the effect of the initial cluster morphology. As a technical check and to ensure no systematic error is introduced by using \texttt{Nemesis}, we also perform runs comparing \texttt{Nemesis} with the direct $N$-body code \texttt{Ph4}. We find no statistically significant differences between algorithms (Appendix~A).
            
            Overall, the final binary fraction from Model~E is $f_{\rm Bin}=0.013$, consistent with \citet{2010MNRAS.404.1835K}, who investigated clusters spanning densities of $\rho = 1-10^{5}$~M$_\odot$~pc$^{-3}$. This indicates that the enhanced binary formation found here arises from the sub-virial, sub-structured initial conditions.
             
            Binaries formed in the Plummer simulations are systematically wider (Figure~\ref{fig:ModelA_vs_ModelC}), reflecting the dynamically relaxed nature of a virialised Plummer sphere. In contrast, sub-virial fractal clusters undergo violent relaxation, producing frequent close encounters that facilitate binary formation and allow tighter orbits. Additionally, stars part of the same clump in fractal models have correlated velocities, which encourages dynamical binary formation \citep[see also,][]{2023MNRAS.525.2907P}. This feature of correlated velocities in local regions agrees well with observations \citep{2013A&A...554A..55H, 2016MNRAS.456.2900A}.

            Nevertheless, our derived rates remain larger than observational measurements (Section~\ref{Sec:Wide_Bin_Forms}). Not accounting for the near-unity binary fraction of protostellar systems \citep{2008AJ....135.2526C, 2013ApJ...768..110C} makes the difference even starker, though difficult to quantify. The discrepancy is partly due to the relatively short integration times, which do not account for the eventual decay of some of the wider systems. Additionally, as we've seen, the sub-virial initial conditions considered here encourage stellar interactions, enabling the formation of dynamical binaries, but do not reflect the typical star cluster configuration (Section~\ref{Sec:Methods_Stars_Planets}). 
            
            Another reason we find a larger binary fraction is that, simulations \citep[i.e;][]{2011MNRAS.418.2565P} and observations \citep[i.e;][]{2012A&A...540A..16M, 2023ApJ...955..134R} indicate that denser clusters exhibit a lower binary fraction. Following this, our low-$N$ cluster likely enhances binary formation. This can also be seen by considering Equation~\ref{Eqn:Eb} and noting that a binary is stable to external perturbations if its binding energy exceeds the local kinetic energy, which translates to $a_{\rm hard}\propto\sigma^{-2}$. Meanwhile, assuming a virialised cluster, $\sigma^{2}\propto M/R\propto N^{2/3}$, leading to $a_{\rm hard}\propto N^{-2/3}$.
            
            Finally, simulations here assume all stars are formed at $t=0$~yr. In reality, a star cluster assembles over $\sim1$~Myr timescales \citep{2011MNRAS.418.1948J, 2012MNRAS.420.1457H, 2023MNRAS.520.5331W}. In their work, \citet{2017ApJ...838..116F, 2019MNRAS.483.4999F} find that a gradual star formation epoch results in the environment remaining in a dense state for longer. Regarding the binary population, this results in more processing of the population, leading to more ionisations. This appears to be particularly important for small-$N$ clusters and binaries with semi-major axis $10^{2}\lesssim a$~[au]$\lesssim10^{3}$ \citep{2023MNRAS.523.2083F}.

            To summarise, both a fractal distribution and Plummer sphere represent idealised initial conditions when modelling star clusters. The former more efficiently produces binaries compared to the latter, with an initially sub-virial state and/or sparser clusters increasing the production efficiency and survival fraction. However, in reality, clusters grow hierarchically and with an extended epoch of star formation \citep[i.e,][]{2021PASJ...73.1036H, 2023ApJ...944..211L, 2024ApJ...975..207C, 2024A&A...690A.207P}. As such, while Plummer models are overly inert, fractal models can be overly violent \citep[see i.e,][]{2011MNRAS.418.2565P, 2024ScPA....3....1P, 2024ApJ...977..203C}.
            
            \subsubsection{Semi-major-axis distributions}
            \citet{2010MNRAS.404.1835K} and \citet{2011MNRAS.415.1179M} both report a bimodal semi-major axis distribution for dynamically formed binaries. The low-$a$ peak arises from dynamical interactions, while the high-$a$ peak originates during cluster dissolution. Comparing their virialised Plummer runs with our own, \citet{2010MNRAS.404.1835K} find peaks at $a\approx10$~au and $a\approx4\times10^{4}$~au (when ignoring extremely wide, unstable system) while we find peaks at $a\approx10^{3}$ and $a\approx3\times10^{4}$~au. The discrepancy in the dynamical peak is due to our cluster being $\sim200$ times less dense at initialisation, which naturally shifts the median semi-major axis to larger $a$ (by approximately a factor $200$).
            
            We do not observe a bimodal distribution in $a$ during fractal runs, with results here more like the observed distribution reported in \citet{2021A&A...649A...6G}. Figure 8 of \citet{2010MNRAS.404.1835K} shows how bimodality in fractal runs becomes more apparent the higher the cluster virial ratio. Combining their results with ours, the erasure of substructures through dynamics allows for a smoothing of the distribution in $a$ in dynamically formed binaries \citep[see also,][]{2021MNRAS.507.4507B}.

        \subsection{Identifying past collisions and captures}
            Several observational features can emerge when planets collide with a star. Tidally stripped planets will produce optical and x-ray transients \citep{2012MNRAS.425.2778M}, and over longer timescales, the accretion of material renders the star more metal rich \citep{1997ApJ...491L..51L, 2001ApJ...556L..59P, 2002ApJ...572.1012S, 2015A&A...582L...6S}. This metallicity excess will dilute over gigayear timescales due to internal mixing. 
            
            Using stellar metallicities, \citet{2021NatAs...5.1163S} estimate that between $20\%-35\%$ of Solar-like stars have experienced a planet collision event. While several binary components have been observed with significant abundance differences relative to their binary partners \citep[e.g.][]{2014ApJ...790L..25T,2015ApJ...801L..10T,2018ApJ...854..138O}, it isn't clear whether this is a result of the highly varied concentration of ice abundances present within a single cloud \citep{2006A&A...453L..47P, 2017MNRAS.467.4753N, 2025NatAs...9..883S}.

        \begin{figure}
            \centering
            \includegraphics[width=\columnwidth]{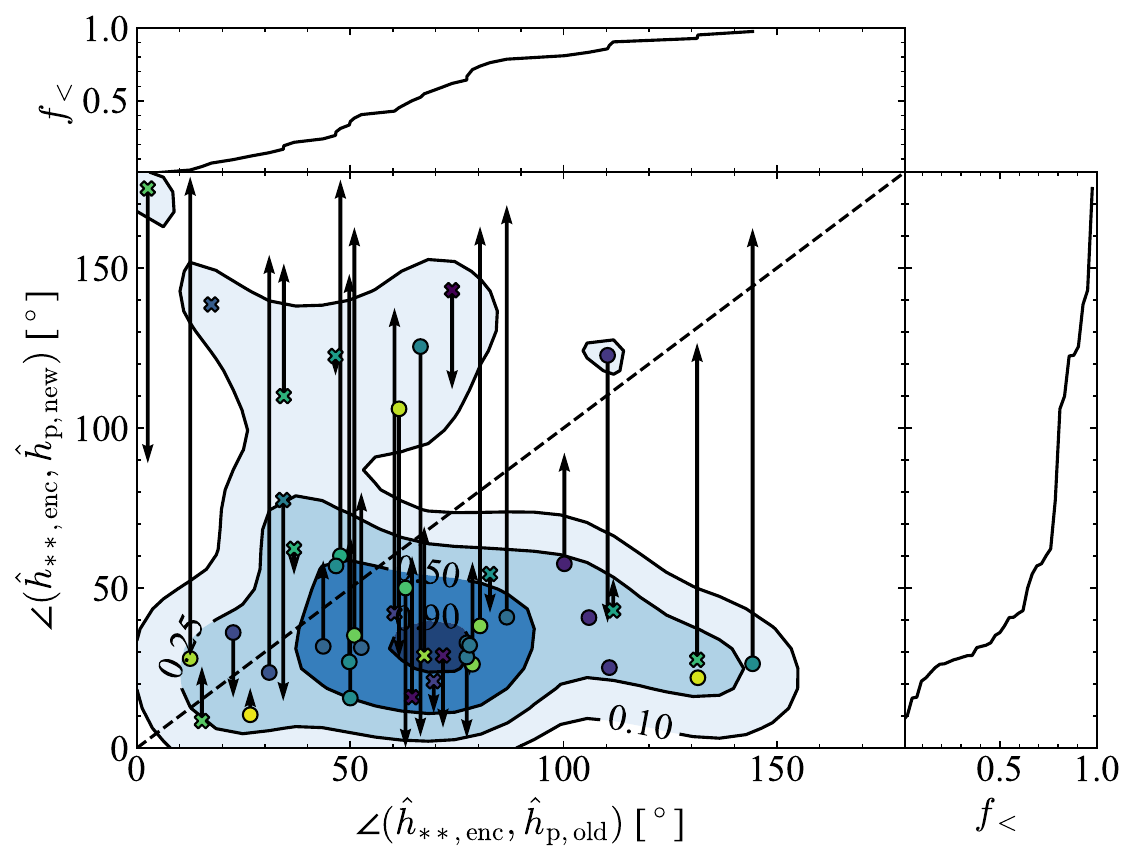}
            \caption{Relative angle between the angular momentum vector of a captured planet and the stellar encounter plane for Model~A. The $x$-axis denotes the relative angle prior to the moment the planet was identified as captured (with resolution $t - t_{\rm cap} \leq 100$~yr). The $y$-axis shows the relative angle at the moment of capture. Arrows indicate the relative angle between the planet's final angular momentum vector and the encounter plane. Crosses denote captures occurring in non-binary systems, while circles denote captures in systems which were identified as binaries for at least $0.1$~Myr prior to capture. Colours of the arrows denote the final semi-major axis. Marker colours are mapped to the particle ID of the capturing star. Side panels show the cumulative distribution function for either relative angle.}
            \label{Fig:System_Inc}
        \end{figure}
        
            Focusing on capture events, we remark that wide orbits can preserve information about the geometry of the most recent close encounter, albeit with some spread. Figure 1 of \citet{2015MNRAS.453.3157J} shows that captured planetesimals can exhibit inclination spreads of up to $\pm30^\circ$ when the closest-approach distance $r_p$ is comparable to the captured particle’s original semi-major axis $a$, with the spread decreasing as $r_p/a$ increases.
            
            In our simulations, capture events span $0.1 \lesssim r_p/a \lesssim 10$, and a similar spread is found between the orbital inclination of captured bodies\footnote{For clips: https://www.youtube.com/watch?v=lxHVny-7ya8} and the encounter plane (see Figure~\ref{Fig:System_Inc}). The $x$-axis corresponds to the angle between the encounter plane and a planet’s orbital angular momentum vector prior to capture. The $y$-axis shows the corresponding angle after capture. The initial orientations are broadly distributed, particularly between $0^\circ$ and $90^\circ$ (median $\langle\angle(\hat{h}_\mathrm{**, enc}, \hat{h}_\mathrm{p, old})\rangle=63.7^{\circ\,\,+16.3}_{\,\,\,\,-19.4}$), in agreement with previous literature showing that capture events occur preferentially during prograde encounters \citep{1993MNRAS.261..190C,2005A&A...437..967P,2023MNRAS.526.1987F}. In contrast, post-capture orientations are concentrated at small angles ($\langle\angle(\hat{h}_\mathrm{**, enc}, \hat{h}_\mathrm{p, new})\rangle=35.6^{\circ\,\,+23.8}_{\,\,\,\,-8.6}$), suggesting that captured planets tend to become partially aligned with the encounter plane, albeit with some spread, consistent with \citet{2015MNRAS.453.3157J}.
            
            The arrows in Figure~\ref{Fig:System_Inc} highlight that, although capture events initially reflect the geometry of the stellar encounter, subsequent dynamical evolution alters the orbital planes of captured planets, since the figure shows the final angle of the captured planet’s orbit. As such, while the encounter geometry may be frozen if it occured in sparse environments \citep[e.g.,][]{2018MNRAS.474.5114C}, in cluster environments, where stellar perturbations are common, this memory of the encounter can get erased over time.

        \subsection{Dynamical vs. Primordial}\label{Sec:Dynamical}
            There is no definitive observation to identify whether a binary has a dynamical or primordial origin. Although dynamical binaries are expected to follow a thermal eccentricity distribution, primordial binaries can also attain large eccentricities \citep[$\langle e\rangle\approx0.4$;][]{2009MNRAS.392..590B,2012MNRAS.419.3115B}. This signal is further obscured by subsequent dynamical evolution. Additionally, the range in semi-major axis overlaps. Fragmentation models typically produce primordial binaries with $a\lesssim10^{2}$~au \citep[$\langle a\rangle\approx20$~au;][]{2009MNRAS.392..590B,2012MNRAS.419.3115B}, while turbulent core fragmentation may lead to binaries with $10^{2}\lesssim a$ [au] $\lesssim10^{3}$~au \citep{2002ApJ...576..870P,2010ApJ...725.1485O}. This latter range accounts for $\sim30\%$ of our binaries (Figure~\ref{fig:ModelA_vs_ModelC}).

            Angular-momentum alignment may provide a more useful, though still imperfect, diagnostic. Primordial binaries might preserve aligned spin--spin, spin--disk, or disk--orbit orientations, whereas dynamically assembled binaries should be closer to isotropic. However, this picture is complicated by accretion, turbulence, and secular evolution. Competitive accretion can align stellar spins in close binaries ($a\lesssim30$~au), but this alignment weakens at larger separations where turbulence dominates \citep{2015MNRAS.450.3306F, 2016ApJ...827L..11O, 2016ApJ...818...73T, 2018MNRAS.475.5618B}. This remains a topic of ongoing research however, as numerical simulations in \citet{2023MNRAS.518.4693G} find spin--spin alignment across primary masses even in turbulent core fragmentation. Likewise, spin-disk orientations will evolve in time \citep{2012Natur.491..418B,2017MNRAS.467.3066A,Albrecht_2022,2026A&A...708A.256V}. Consequently, the relative orientation of orbital planes within multiple systems may provide a more promising diagnostic.
            
            Previous numerical work have found that binaries formed with separations $a\lesssim100$~au tend to have their orbital plane aligned with any associated disks or planetary systems \citep{2012MNRAS.419.3115B,2018MNRAS.475.5618B}. Alignment may even occur in wider systems through dissipative precession \citep{2024ApJ...972..161G}. Observations have shown evidence for such alignment to exist in young systems, for example among Herbig Ae/Be binaries \citep{2006MNRAS.367..737B,2011A&A...532A..28W} where alignment is considerable at $40\lesssim a$ [au] $\lesssim 200$.
            
            \begin{figure*}
                \centering                
                \includegraphics[width=\textwidth]{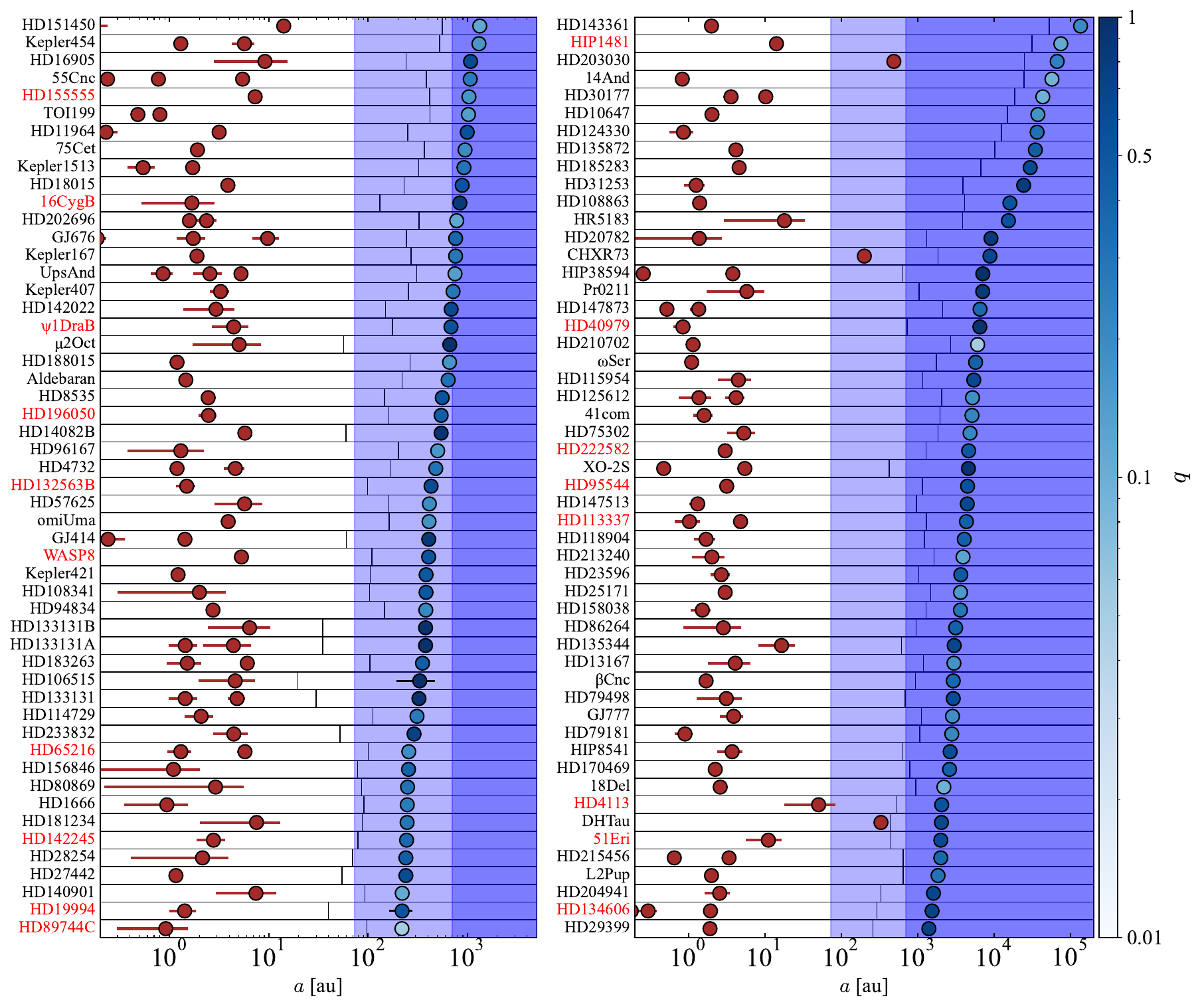}
                \caption{Same as Figure~\ref{Fig:ModelA_Overview} but for observational data \citep[taken from][]{2025A&A...700A.106T}. Here, only systems with $a_{\rm bin}>200$~au which include planetary orbits $a>0.2$~au are considered. Binary eccentricities are generally not measured and are therefore assumed to be $e=0$, this is also the case for a subset of the planetary orbits. Names in red are part of a triple or higher-order system. The light-shaded region spans $a \geq 73$~au. Its lower bound corresponds to the transition from predominantly aligned planetary–binary orbital architectures to an isotropic inclination distribution \citep{2025arXiv250925332Z} assuming $e=0.45$. The darker shaded region spans $a$~au $\gtrsim700$. This lower bound reflects the more conservative transition between coplanar and isotropic systems identified by \citet{2022AJ....163..207C, 2025AJ....169..308C}.}
                \label{Fig:ObsBinary}
            \end{figure*} 
            Notably, recent observations reveal a dichotomy in the relative inclination distribution between planetary systems and their host binaries. Systems with projected separations $\rho\gtrsim700$~au tend toward isotropic planet-binary inclinations, whereas tighter binaries are preferentially aligned \citep{2022AJ....163..207C,2025AJ....169..308C,2023AJ....166..166L}. \citet{2025arXiv250925332Z} suggest that this transition corresponds to periastron distances of $r_p\approx40$~au. Comparable trends are also seen in hierarchical triples, where alignment between inner and outer orbital planes increases at smaller separations \citep{2025PASP..137i4201S,2026ApJ...998..151T}. In combination with our results (Table~\ref{Tab:BinStats}) and assuming that a vast majority ($\lesssim95\%$) of primordial binaries form with $a<10^{3}$~au \citep{2009MNRAS.392..590B,2012MNRAS.419.3115B}, this dichotomy in relative alignments may reflect a transition between dominant binary formation channels. We shade these regions for the observed systems shown in Figure~\ref{Fig:ObsBinary}.
            
            Mass ratio of stellar components can provide an additional indicator. Namely, dynamical binaries tend towards a near-flat mass-ratio distribution while primordial binaries have this behaviour restricted to systems with $a\lesssim10$~au \citep{1979AJ.....84..401L, 2000A&A...360..997T, 2009MNRAS.392..590B, 2012MNRAS.419.3115B}. Since dynamical binaries tend to form wide (table~\ref{Tab:BinStats}), this can be used as a first-order indicator. Greater confidence can be achieved for very wide systems. 
        
        \subsection{Forming S-Type systems}\label{Sec:SType_Disc}  
            The planet occupancy around binaries become statistically indistinguishable from those around single stars when the binary semi-major axis $a > 1500$~au \citep{2014ApJ...791..111W, 2016AJ....152....8K, 2020Galax...8...16B}. \citet{2026A&A...708A..37V} argue this is the case down to $a\gtrsim10^{2}$~au. 
            
            \begin{table}
                \caption{Median and interquartile spread of ejection velocities for rogue planets in Model~A and Model~B. Top row corresponds to rogue planets ejected while part of a binary system, the bottom row those ejected while being part of an isolated stellar system.}
                \label{Tab:AlphaCen} 
                \centering 
                    \begin{tabular}{c c c}
                    \hline\hline
                    & Model~A & Model~B \\
                    \hline \vspace{-1em}\\
                    Isolated & $0.97^{+1.69}_{-0.67}$~km~s$^{-1}$ & $0.65^{+0.87}_{-0.37}$~km~s$^{-1}$ \vspace{0.1em} \\
                    Binary & $2.22^{+2.77}_{-1.59}$~km~s$^{-1}$ & $1.41^{+0.26}_{-0.99}$~km~s$^{-1}$ \vspace{0.1em} \\ \hline
                \end{tabular}
            \end{table}
            For Model~A, we find that binary S-type systems become prevalent when stellar components have semi-major axis $a\gtrsim800$~au. As of writing, there are eight confirmed binary S-type systems observed. Additionally, HD 135344 is a binary whose stellar components have projected separations $\rho\approx2800$~au and contains a planet orbiting the primary while the secondary hosts a misaligned protoplanetary disk \citep{2025A&A...700A..21S}. Our results indicate that many more should exist  (Figure~\ref{Fig:ObsBinary}), particularly those shaded in dark blue, should be expected to be binary S-types. Additionally, we find that the presence of planets within a system is rare if the stellar components have $r_p<100$~au, a result agreeing with observations which instead find a $a<100$~au threshold \citep{2007A&A...462..345D, 2011IAUS..276..409E, 2021MNRAS.507.3593M}. 
            
            Interestingly, our simulations indicate a difference in ejection velocities for rogue planets once part of a isolated system, or once part of a binary. While the distributions overlap, the difference is stark and measurements of relatively quick rogue planets in young environments can indicate a previous binary-origin.

        \subsection{Lack of P-Type systems}\label{Sec:PType_Disc}
            Our simulation's failure to produce P-type systems suggests that this architecture is unlikely to arise through the late dynamical assembly of a star cluster, and are more likely primordial. Interestingly, six of the ten known planetary-mass companions with projected separations $\gtrsim350$~au orbit stellar multiple systems \citep{2010ApJ...725.1405B, 2014ApJ...780L...4B, 2018AJ....156...57D, 2023MNRAS.519.1688D, 2021Natur.600..231J, 2024AJ....167..253R, 2025A&A...704A.221V}. The observed stellar binaries are considerably more compact than the dynamically formed binaries found in our simulations. The widest (known) orbit has projected separation $\gtrsim10$~au \citep{2025A&A...704A.221V}, which remains several times smaller than even the tightest binary formed here. 
            
            An in-situ formation of the planet at these wide orbits is complicated due to migration. Across different-sized systems, \citet{2011MNRAS.416.1971B} find that a Saturn-mass body will migrate from $\sim10^{2}$~au to $20$~au within $\sim5\times10^{3}$~yr. This timescale rarely exceeds $\lesssim10^{5}$~yr, with the inspiral timescale decreasing with increasing planet mass. This leaves open the possibility that some wide companions were captured or formed through a non-standard channel. More observations are needed to distinguish between these scenarios. 

            Regarding the former, binaries typically have a greater capture cross-section than isolated stars. This is due to their larger geometric cross-section and their greater mass (generally). However, for the distances measured, the former aspect is essentially negligible as the stellar multiple acts as a point-mass for the wide orbit planet. 
            
            Finally, the classification of directly imaged objects requires caution as inferred masses are sensitive to atmospheric and evolutionary models. Recently, \citet{2026A&A...710A.328D} included dust extinction and circumplanetary-disc emission, finding that the spectra of YSES~1b is more like a Brown Dwarf, with mass $25.7^{+4.1}_{-3.6}$~M$_{\rm Jup}$ rather than a Jupiter-mass object with mass $14\pm3$~M$_{\rm Jup}$ \citep[see also,][ for a similar relabelling of planetary-mass objects to Brown Dwarf]{2025MNRAS.542L.126L}. This distinction has little influence on the dynamics, but is relevant since numerical work on star formation has shown that multiple substellar fragments can emerge during disc fragmentation. Over time, gravitational interaction between the fragments will lead to ejection of the lower-mass fragments, or scatter them onto wide orbits, while the more massive survivors remain bound on a tight, yet eccentric orbit \citep{2002MNRAS.336..705B, 2003MNRAS.346L..36R}. As such, while some P-type systems may be captured, depending on their masses, their origin may also be more closely related to the formation of brown dwarfs than planetary systems.
        
        \subsection{Hierarchical triples and the case for $\alpha$Centauri}\label{Sec:Triples}
            Across Model~A runs, $24$ triple systems are present at $t=30$~Myr. This corresponds to a triple formation (and survival) rate of $f_{\rm T}=0.02$, consistent with the $f_{\rm T}=0.005-0.03$ formation rate found in \citet{2019MNRAS.483..648F}. Of these systems, $87\%$ of the outer companions host a planetary system. In contrast, only $13\%$ of inner primaries and $52\%$ of inner secondaries retain planets. This low survival rate within the inner system is due to triples tending to have a comparatively tight inner binary ($\langle a\rangle=527^{+776}_{-329}$~au) relative to the outer companion ($\langle a\rangle=4.21^{+2.56}_{-3.00}\times10^{4}$~au). Among the triples, $5/24$ have their outer-binary orbit with $a<10^{4}$~au. Of these, two have at least one of the inner members retain a planet. These systems have inner semi-major axes of $a_{\rm in}=87$~au and $a_{\rm in}=820$~au, and are considered analogous to $\alpha$Centauri, although the latter is substantially wider.

            \begin{table}
                \caption{Orbital parameters for $\alpha$Cen A and $\alpha$Cen B (top row) and of the inner binary with the outer companion, $\alpha$Cen C. Values are taken from table 8 and 10 of \citet{2021AJ....162...14A}.}
                \label{Tab:AlphaCen} 
                \centering 
                    \begin{tabular}{c c c c}
                    \hline\hline
                    & $a$ [au] & $e$ & $i$ [$^{\circ}$] \\
                    \hline \vspace{-0.75em}\\
                    A-B & $23.30\pm0.01$ & $0.519$ & $79.2$ \\
                    (AB)-C & $8.2\times10^{3}$ & $0.497$ & $124.9$ \\\hline
                \end{tabular}
            \end{table}
            
            Being our nearest cosmic neighbour, the $\alpha$Centauri system has long inspired astronomers, even appearing in Ptolemy's Almagest in the second century. The orbital parameters of the system is given in Table~\ref{Tab:AlphaCen} and has component masses $m_*=1.08$~M$_\odot$ ($\alpha$Cen A), $m_*=0.909$~M$_\odot$ ($\alpha$Cen B) and $m_*=0.122$~M$_\odot$ ($\alpha$Cen C) \citep{2021AJ....162...14A, 2017A&A...598L...7K}. $\alpha$Cen A has a candidate planet of mass $m_p=120\pm30$~M$_\oplus$ orbiting with $e=0.4$ and $a\approx2$~au \citep{2021NatCo..12..922W, 2025ApJ...989L..23S, 2025ApJ...989L..22B}, $\alpha$Cen B hosts no known planets while $\alpha$Cen C has two confirmed planets \citep{2016Natur.536..437A, 2020A&A...639A..77S, 2025A&A...700A..11S, 2022A&A...658A.115F}, and a disputed third planet \citep{2020SciA....6.7467D, 2022AJ....164...84A}.
            
            Previous numerical work has shown that if the current orbital parameters reflect their birth parameters, planet growth beyond $\approx0.50-0.75$~au is strongly suppressed due to dynamical stirring of planetesimals hindering accretion \citep{2008MNRAS.388.1528T}. To explain the observed location of the planet, the authors mention outward migration although they note this remains unlikely since gas disc driven migration is generally inward \citep[see also][]{2007prpl.conf..655P}, while outward planetesimal-driven migration would require conditions not seen for terrestrial planets at such small radii. Alternatively, \citet{2008MNRAS.388.1528T} mention that the inner binary may have been initially wider and less eccentric.

            Within the truncation-only planetary population-synthesis model of \citet{2026A&A...708A..37V}, the authors found that planets with mass $m_p\gtrsim 100$~M$_\oplus$ struggled to form in a binary composed of a primary with mass $1$~M$_\odot$ and a secondary with mass between $0.1 - 1$~M$_\odot$ irrespective of the binary mass ratio and eccentricity as long as the binary semi-major axis was $a<40.6$~au. This threshold increases when accounting for the gravitational perturbations induced by the secondary, the stirring up of eccentricities and reducing the accretion efficiency \citep{2026A&A...708A..38N}, making a dynamical origin even more viable.
            
            Even so, an initially wider $\alpha$Cen system into its current configuration can be explained by the unfolding of an initially compact, unstable triple. Here, a compact triple has the tertiary scatter onto a wide orbit since it absorbs the energy from the inner stellar binary. This removal of energy consequently hardens the inner binary \citep{2012Natur.492..221R}. This mechanism would predict a slightly higher eccentricity ($e\approx0.7$, see their figure~2) for the tertiary relative to the inner binary's centre of mass than observed but nevertheless remains compatible.

            Here, we show another possibility; one of a purely dynamical origin whose delay time in binary (and triple) formation does not hinder planet formation. This interpretation is also motivated by the large inclination offset between the candidate planet around $\alpha$ Cen A \citep[$50^{\circ}$ or $180^{\circ}$,][]{2025ApJ...989L..22B} with the orbital plane of the inner binary. As discussed in Section~\ref{Sec:Dynamical}, numerical and observational studies generally find alignment in tight systems ($a\lesssim200$~au), although notable exceptions do exist \citep[e.g.][albeit without knowing their formation channel]{2014Natur.511..567J}. Given the relatively small number of simulations performed here, the emergence of analogous systems is particularly striking. 

    \section{Limitations}\label{Sec:Limitations}
        Numerical simulations will always require assumptions in their initial conditions, and this is no different here. Below are several assumptions or omissions made in our numerical work and their impact on results.
        \subsection{Lack of primordial binaries} 
            The simulations contain no primordial binaries, which allows us to better isolate the dynamically assembled population. While primordial binaries have little influence on the cluster as a whole while gas is present \citep{2023MNRAS.521.1338C, 2026arXiv260810750H}, they do become important later on as they act as a heat source to expand the cluster \citep{1987degc.book.....S}. 
            
            Additionally, primordial binaries alter the underlying properties of the emerging dynamically-formed binary population. For instance, figure~13 of \citet{2021MNRAS.501.4464C} indicates that in the presence of primordial binaries, the primary mass, semi-major axis and eccentricity of dynamically formed binaries is smaller than in clusters with no primordial binaries. In their numerical work \citet{2010MNRAS.404.1835K} find that the presence of primordial binaries reduce the wide binary fraction. In essence, the presence of primordial binaries will reduce the number of wide binary systems. This makes it especially relevant here since planetary stability depends strongly on the stellar binary orbit, and namely its semi-major axis \citep[see Figure~\ref{Fig:ModelA_Overview} and][]{1999AJ....117..621H}. As a result, a shift towards more compact binaries may mean some of the wider planetary orbits observed here would not survive. 
            
            We note that primordial binaries are also likely to more efficiently ionise systems due to their greater interaction cross-section and that their omission results in us likely not finding many P-type systems (Section~\ref{Sec:Dynamical}). Beyond these, the presence of primordial binaries likely results in a richness of emergent phenomena and we leave it for future work to consider them.
            
        \subsection{Idealised Circumstellar Systems}
            The planetary systems in our simulations are initially fully formed, their orbits are initially dynamically cold and stable, and no circumstellar gas or disk is present. Gas within circumstellar disk has a lifetime of $\sim10$~Myr \citep{2022ApJ...939L..10P,2026ApJ..1003..119P}. As a result, if binary stars assemble before this dissipation timescale, remaining gas will damp any eccentricity and inclination excitation \citep{2014prpl.conf..667B,2023MNRAS.526.1987F}. This has several effects; it reduces the formation of orbit-crossing configurations, it can drive migration, potentially moving planets onto more tightly bound orbits that are less susceptible to subsequent stellar encounters. By omitting these processes, our models may overestimate the volatility of planetary systems and, consequently, the frequency of planet ejections.
            
            Encounters may truncate, strip, or misalign protoplanetary disks. These interactions can inhibit planet formation by depleting the gas and dust reservoir, altering the masses and architectures of the systems that eventually emerge. This too, can reduce system volatility as lower-mass embryos exert weaker gravitational effects onto one another, leading to fewer ejections. The presence of gas can also influence the stellar population itself, as it will facilitate the migration of binaries leading to more compact systems \citep{1995MNRAS.277..362B,1998MNRAS.300..292K,2012ApJ...750...34H,2018MNRAS.475.5618B}. Its omission may therefore also contribute to the scarcity of compact binaries in our simulations.
            
            Finally, the innermost planets in our initial systems have orbital periods $P>1$~yr. In contrast, only $10.1\%$ of the planets listed in the Extrasolar Planets Encyclopaedia have $P>1$~yr\footnote{\url{https://exoplanet.eu/catalog/}}. Although this fraction is strongly affected by observational bias (survey times and selection functions of transit and radial-velocity surveys), close-in planets are intrinsically more resistant to perturbations from a stellar companion \citep{1999AJ....117..621H} and including them would therefore likely increase the number of planets retained in binary S-type configurations.

        \subsection{Finite integration time}
            The simulations here are integrated until $30$~Myr. By this time, the cluster has essentially dissolved (Figure~\ref{Fig:cluster_density}). Assuming a solar neighbourhood-like environment, we do not expect binaries to have their parameters modified, even over a Hubble time \cite{1975AJ.....80..809H}. The situation is different for the planet-mass objects part of the binary. Perturbations can seed instabilities and planet ejections. Chaos makes this difficult to predict, with secular effects potentially accumulating over over gigayear long timescales \citep{2019MNRAS.488.1366L, 2026arXiv260903575P}. Using the empirical results of previous work \citep{1999AJ....117..621H, 2020AJ....159...80Q}, we expect planets part of a binary with $a\gtrsim 10^{3}$~au or planets with a semimajor-axis $<8\%$ the binary's semi-major axis to be intrinsically stable. While this constitutes most of the systems, particularly if we account for many planets having tight orbits, it does imply that wide planetary-mass objects should become increasingly rare within binary systems over time.
            
\section{Conclusions}
    The paper here builds on previous literature, and considers the evolution of planetary systems orbiting around dynamically formed binaries within a star cluster environment until its dispersion. Below are some of our main findings:
    \begin{enumerate}
        \item There is a disproportionately high number of collision (Section~\ref{Sec:CollEvents}) and planet capture events (Section~\ref{Sec:CaptEvents}) occurring around binary stars. A vast majority of collisions occur in systems that is, or has once been part of a binary, with planet-planet collisions more common in systems initially not in mean motion resonance (Section~\ref{Sec:System_Architecture}).
        \item Sub-clustered environments are more likely to create dynamically formed binaries than a virialised isotropic one (Section~\ref{Sec:Env}), many of which are binary S-type systems. This is also reflected in the number of very wide binaries ($a>10^{4}$~au).
        \item For the cluster configuration considered here, binary S-type systems are especially prevalent (Section~\ref{Sec:SType_Disc}). Contrastingly, it appears difficult to dynamically form P-type systems (Section~\ref{Sec:PType_Disc}).
        \item Rogue planets originating from a binary system escape at greater velocities than those originally orbiting an isolated star (Section~\ref{Sec:SType_Disc}).
        \item Given the observed orbital parameters of planets within the $\alpha$Centauri system, we propose that our nearest cosmic neighbour may have formed dynamically (Section~\ref{Sec:Triples}).
    \end{enumerate}
    There are many ways to extend this work. Past work has already indicated that super-virial systems and primordial binaries will incur fewer dynamically formed binaries. Presence of the latter population could also explain P-type systems, and analysis of their evolution during cluster evolution would be insightful both in seeing how circumbinary planets evolve, but also investigating the rates of captured planets around such systems.
        

\section{Energy consumption}
    Due to \texttt{Nemesis}' fluctuating CPU usage, it is difficult to estimate the actual power outage. Assuming that $7$ cores are running at $100\%$ power throughout run time. and given the total time spent on simulation was $2100$ hours, we get a final wall clock time of $14700$~hr. Intel Xeon Gold 5220R processors were used, each being powered by $150$~W. The total energy consumed is thus $E\approx 14700\times150\times\frac{1}{24}=91.875$~kWh. 
    
    In 2025, the Netherlands had a CO$_2$ equivalent of $0.388$~kg~kWh$^{-1}$, of which $28\%$ of the energy produced was renewable\footnote{https://www.nowtricity.com/country/netherlands/}. Factoring this green energy, the kWH to CO$_2$ equivalent emission is $0.27936$~kg~kWh$^{-1}$. As such, the total equivalent CO$_2$ emission is $E_{\rm CO_2}\approx (91.875\, {\rm kWh})\times(0.27936 {\rm\, kg\, kWh^{-1}})\approx25.7$~kg. This roughly amounts to $12850$ images generated by DALL-E \citep{2024NatSR..14.3732T} or roughly three to four servings of beef or about $200$ servings of rice \citep{2018Sci...360..987P}.

\begin{acknowledgements}
    Analysis was made using the open-source \texttt{Python} packages \texttt{NumPy} \citep{2020Natur.585..357H} and \texttt{Matplotlib} \citep{2007CSE.....9...90H}. We would like to thank the anonymous referee whose suggestions helped improve the paper.
\end{acknowledgements}

\bibliographystyle{aa}
\bibliography{references.bib}

\begin{appendix}
    \section{Direct $N$-Body versus \texttt{Nemesis}}\label{Sec:App}
        \begin{table}[h!]
            \caption{Summary of binary properties across ten runs using \texttt{Nemesis} (column 1) and a \texttt{Ph4} (column 2). Row 1: Median and error in the cluster virial radius. Row 2: Median and error in the cluster virial ratio. Row 3: Median number of binary systems. Row 4: Median and error in the binary semi-major axis. Row 5: Median and error in the binary eccentricity.}
            \label{Tab:Appendix}
            \centering 
                \begin{tabular}{c c c}
                \hline\hline
                & \texttt{Nemesis} & \texttt{Ph4} \\
                \hline \vspace{-0.75em}\\
                $\langle R_{\rm vir}\rangle$ [pc] & $0.42\pm0.07$ & $0.31\pm0.04$ \vspace{0.2em}\\
                $\langle Q_{\rm vir}\rangle$ & $0.73\pm0.04$ & $0.69\pm0.04$ \vspace{0.2em}\\
                $\langle N_{\rm bin} \rangle$ & $2.0\pm0.4$ & $2.0\pm0.5$ \vspace{0.2em}\\
                $\langle a \rangle$ [au] & $(2.9\pm0.9)\times10^{4}$ & $(2.7\pm0.8)\times10^{4}$ \vspace{0.2em}\\
                $\langle e \rangle$ & $0.68\pm0.05$ & $0.69\pm0.05$ \vspace{0.2em} \\ \hline
            \end{tabular}
        \end{table}

        To test whether \texttt{Nemesis} artificially amplifies or suppresses the binary formation rate, additional runs are done. Ten independent realisations of a virialised Plummer sphere with virial radius $r_h=0.5$~pc and consisting of $N_*=150$ stars are conducted, with no planetary systems. Each configuration was integrated for $30\,\mathrm{Myr}$. For the direct $N$-body reference, only \texttt{Ph4} was used, while for \texttt{Nemesis} all parameters were kept identical to those described in Section~\ref{Sec:Numerics}. In both cases, the internal time step is fixed to $\eta=2^{-3}$. The results are summarised in Table~\ref{Tab:Appendix}.
        
\begin{figure}
                \centering                
                \includegraphics[width=\columnwidth]{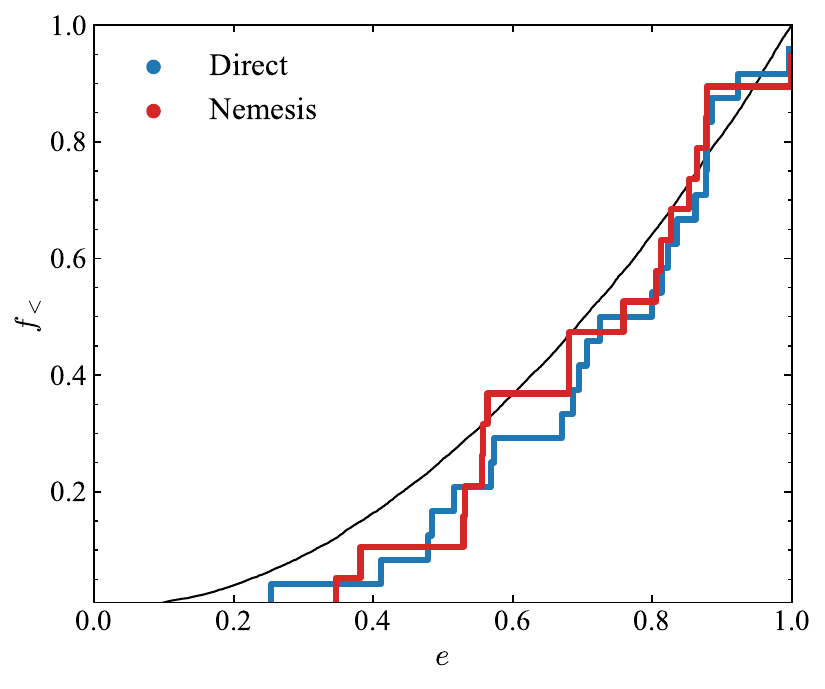}
                \caption{Eccentricity cumulative distribution function of dynamically formed binaries during Plummer runs.}
                \label{Fig:Appendix_ecc}
        \end{figure}

\begin{figure}
                \centering                
                \includegraphics[width=\columnwidth]{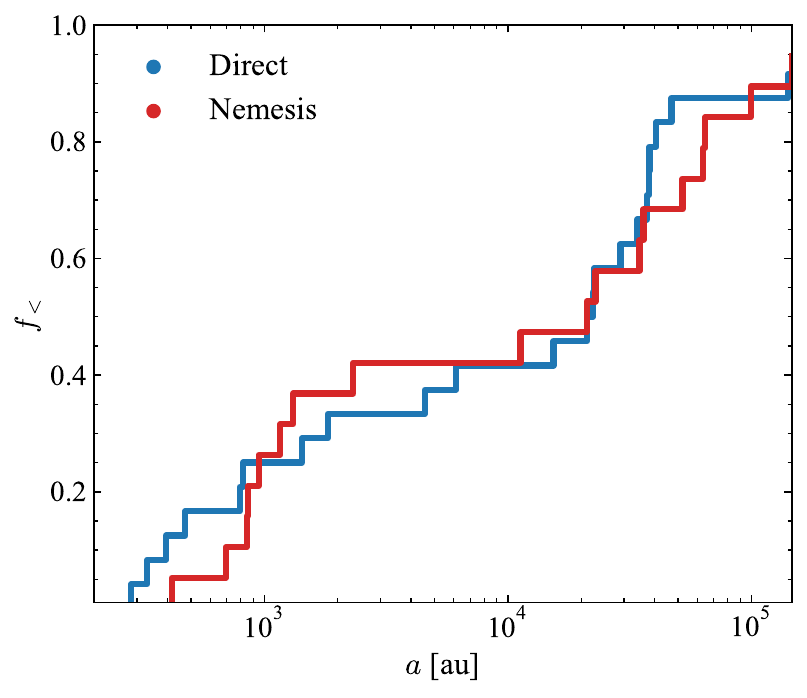}
                \caption{Semi-major axis cumulative distribution function of dynamically formed binaries during Plummer runs.}
                \label{Fig:Appendix_sma}
        \end{figure}

        For all quantities analysed, the results obtained with both algorithms agree within statistical uncertainties. Crucially, both the number of binaries and their orbital parameter distributions are consistent. This agreement is illustrated by the cumulative distribution functions of binary eccentricity and semi-major axis shown in Figure~\ref{Fig:Appendix_ecc} and Figure~\ref{Fig:Appendix_sma} respectively. Using a Kolmogorov-Smirnov test, $p$-values of $p_a=0.81$ and $p_e=0.85$ are obtained, thus we do not reject the null hypothesis that the curves are sampled from different underlying distribution.
        
        The increase in the virial radius for \texttt{Nemesis} could be a byproduct of how it resolves close encounters. Indeed, as mentioned in the main text, binaries act as a heat source for the cluster and interactions with perturbers will lead to cluster expansion. Using a symplectic integrator rather than a direct $N$-body code can lead to deviations. While such integrators are good for long-term evolution of systems, they also fall short in resolving abrupt changes since their internal time-step criteria remains fixed and are not adaptive, like in the case of \texttt{Ph4}. This can lead to poorly resolved close encounters and is something to consider in future investigations. 
        
        Overall, given the intrinsic sensitivity of gravitational systems to initial conditions, some scatter between realisations is expected. The fact that all measured quantities remain consistent between the two algorithms strongly suggests that \texttt{Nemesis} does not introduce any artificial enhancement or suppression of binary formation.
        
        
\end{appendix}

\end{document}